\documentclass[11pt,twoside
]{article}

\usepackage{jae2016} 
\usepackage{graphicx}
\usepackage{subfigure}
\usepackage{psfrag}
\usepackage{amssymb}
\usepackage[spanish,activeacute,english]{babel}
\usepackage[latin1]{inputenc}
\usepackage[T1]{fontenc} 
\usepackage{ae,aecompl} 
\usepackage{latexsym}
\usepackage{verbatim}
\usepackage{amsmath}
\usepackage{stmaryrd}
\usepackage{amsfonts}
\usepackage{amssymb}
\usepackage{wasysym}
\usepackage[colorlinks=true,dvips]{hyperref}
\begin{document}
\myselectenglish
\vskip 1.0cm
\markboth{J. C. Forte}%
{Synchronized Globular Cluster Formation.}

\pagestyle{myheadings}

\title{About Synchronized Globular Cluster Formation over Supra-galactic Scales.}

\author{
Juan C. Forte$^{1,2}$,
}
\affil{%
  (1) Consejo Nacional de Investigaciones Cient\'ificas y T\'ecnicas, Rep. Argentina.\\
  (2) Planetario "Galileo Galilei", Ministerio de Planificaci\'on e Innovaci\'on Tecnol\'ogica,
  Ciudad Aut\'onoma de Buenos Aires.\\
}
\begin{abstract} 
 Observational and theoretical arguments support the idea that violent
 events connected with $AGN$ activity and/or intense star forming episodes
 have played a significant role in the early phases of galaxy formation at
 high red shifts. Being old stellar systems, globular clusters seem
 adequate candidates to search for the eventual signatures that might
 have been left by those energetic phenomena. The analysis of the
 colour distributions of several thousands of globular clusters in the
 Virgo and Fornax galaxy clusters reveals the existence of some 
 interesting and previously undetected features. A simple pattern recognition
 technique, indicates the presence of "colour modulations", distinctive for
 each galaxy cluster.\\
 These patterns were first found on composite samples of globular clusters
 in galaxies with $M_{g}=$-20.2 to -19.2, and later, detected on some
 sub-samples of globular associated with individual giant elliptical galaxies.
 The results suggest that the globular cluster formation process has not
 been completely stochastic but, rather, included a significant fraction
 of globulars that formed in a synchronized way and over supra-galactic 
 spatial scales. A tentative approach indicates that the putative events that
 enhanced globular cluster formation took place during a time lapse of 1.5 $Gy$
 and in a range of red shifts $z$ between 2 and 4.   
\end{abstract}
\section{Introduction.}
 Being old  stellar systems, and potential carriers of valuable information of the
 astrophysical processes that characterized the early Universe, globular clusters
 ($GCs$) have become the target of an increasing volume of research on both the
 observational and theoretical fronts.\\
 Even so, some historical questions are still open. For example, why
 $GC$ formation is not detected in low red shift galaxies even though enough
 interstellar matter is available to fuel such a process ? Which is the
 "missing" ingredient not operating at low red shifts ? Nearby galaxy mergers do
 exhibit some massive clusters that seem to resemble young  globulars and suggest that
 violent/energetic phenomena might be such an ingredient.\\ 
 A key subject in the study of globular cluster systems is the analysis
 of the globular cluster colour distribution ($GCCDs$ in what follows). A
 proper decoding of these distributions imply the understanding of the
 connection between ages, chemical abundances and spatial distributions.\\
 Frequently, these distributions exhibit the so called "colour bimodality", i.e., 
 the presence of two dominant ("blue" and "red") $GC$ populations.\\ 
 Peng et al. (2006) have shown that bi modality is in fact a feature that
 depends on galaxy mass. The red $GC$ population becomes less evident with
 decreasing mass and disappears in the less massive galaxies. In turn, 
 "blue" globulars seem to be present in almost all galaxies.\\
 This characteristic led Cen (2001) to suggest that blue globulars
 formed in a synchronized way and as the result of a very large scale
 phenomenon: The re-ionization of the  Universe. An underlying question 
 is, if other highly energetic events might have left some kind of
 detectable features, for example, on the $GCCDs$.\\
 Fabian (2012) presents a number of observational results that point out
 the important role of $AGN$ activity in producing massive outflows that
 change the environmental conditions of interstellar gas on large spatial
 volumes. On the other side, the most recent models of galaxy formation
 (e.g. Vogelsberger 2014)  include $AGN$ phenomena and lead to
 remarkably realistic results.\\
 In general, these phenomena are expected to produce star forming "quenching".
 However, several results in the literature argue in the other direction,
 i.e., that under given circumstances, these effects end up in enhancing the
 star forming process (see, for example, Silk 2013).\\
 In the particular case of $GCs$, and in a $\it naive$ approach, one might
 expect that quenching/enhancement would be reflected as valleys/peaks in
 the $GCCDs$. So far these kind features have not been reported.\\
 Eventually, this situation may suggest that the usual approach in analysing
 the $GCCDs$ needs a change in the strategy. For example, the composite samples
 of $GCs$ belonging to galaxies with comparable brightness (mass) will
 presumably erase the characteristics of a given globular cluster system but,
 on the other side, might enhance the presence of systemic features common to
 these galaxies.\\
 This type of approach is the core of the following analysis. The details of
 the pattern recognition procedure, as well as other inherent results, can be
 found in Forte (2017; MNRAS, submitted).
\section{Data Sources}
\begin{figure}[!ht]
  \centering
  \includegraphics[width=.5\textwidth]{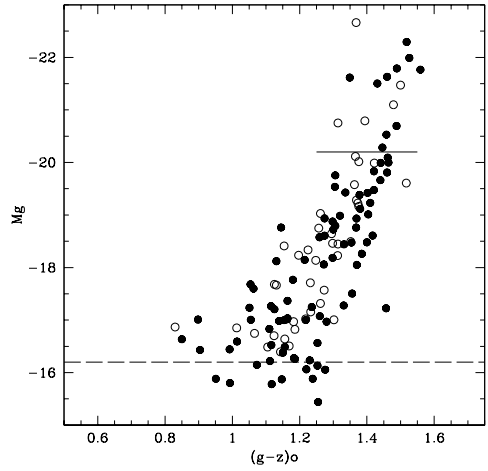}~\hfill
  \caption{Absolute magnitude vs. (g-z) colour for galaxies in the Virgo 
 (filled circles) and Fornax (open circles) $ACSs$. Objects above the horizontal
 line are giant galaxies. The dashed line indicates the faintest limit of
 the analysis.
}
  \label{fig:fig1}
\end{figure}

  The main data source in this work is the $g,z$ photometry for $GCs$ associated with
  galaxies in the Virgo and Fornax clusters presented by Jord\'an et al. (2006, 2015).
  We also revisit the Washington photometry given by Forte, Faifer \& Geisler (2007)
  for the central giant galaxies $NGC~1399$ and $NGC~4486$ and, in the case
  of this galaxy, the $griz$ $Gemini-GMOS$ photometry  by Forte et al.
  (2013).\\
  We adopt a $GCs$ limiting magnitude $g=$25.0 in order to guarantee both a high
  spatial completeness and $(g-z)$ colour errors below $\approx 0.07$ mag. 
  In summary, these data correspond, to 7671 $GCs$ in 88 Virgo galaxies and
  4317 $GCs$ in 42 Fornax galaxies.\\
  Fig.~\ref{fig:fig1} shows the colour magnitude diagram for the Virgo and
  Fornax galaxies. Galaxies above the solid horizontal line are considered as
  giants. The dashed horizontal line is the faintest limit of the analysis.\\
\section{Colour patterns recognition.}
\begin{figure}[!ht]
  \centering
  \includegraphics[width=.5\textwidth]{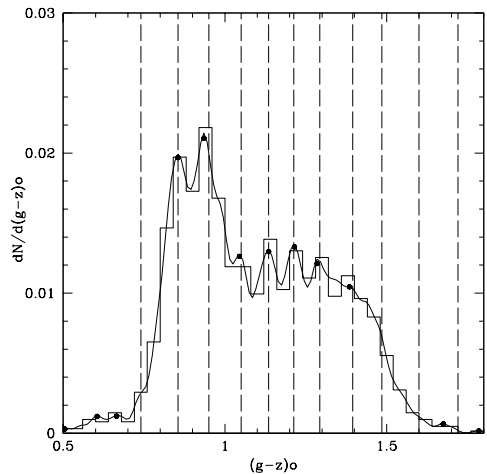}
  \caption{Discrete and smoothed colour distribution for 1531 clusters in 13
Virgo galaxies with $M_{g}=$-20.2 to -19.2. The dashed lines indicate the Virgo
colour pattern.
}
  \label{fig:fig2}
\end{figure}

\begin{figure}[!ht]
  \centering
  \includegraphics[width=.5\textwidth]{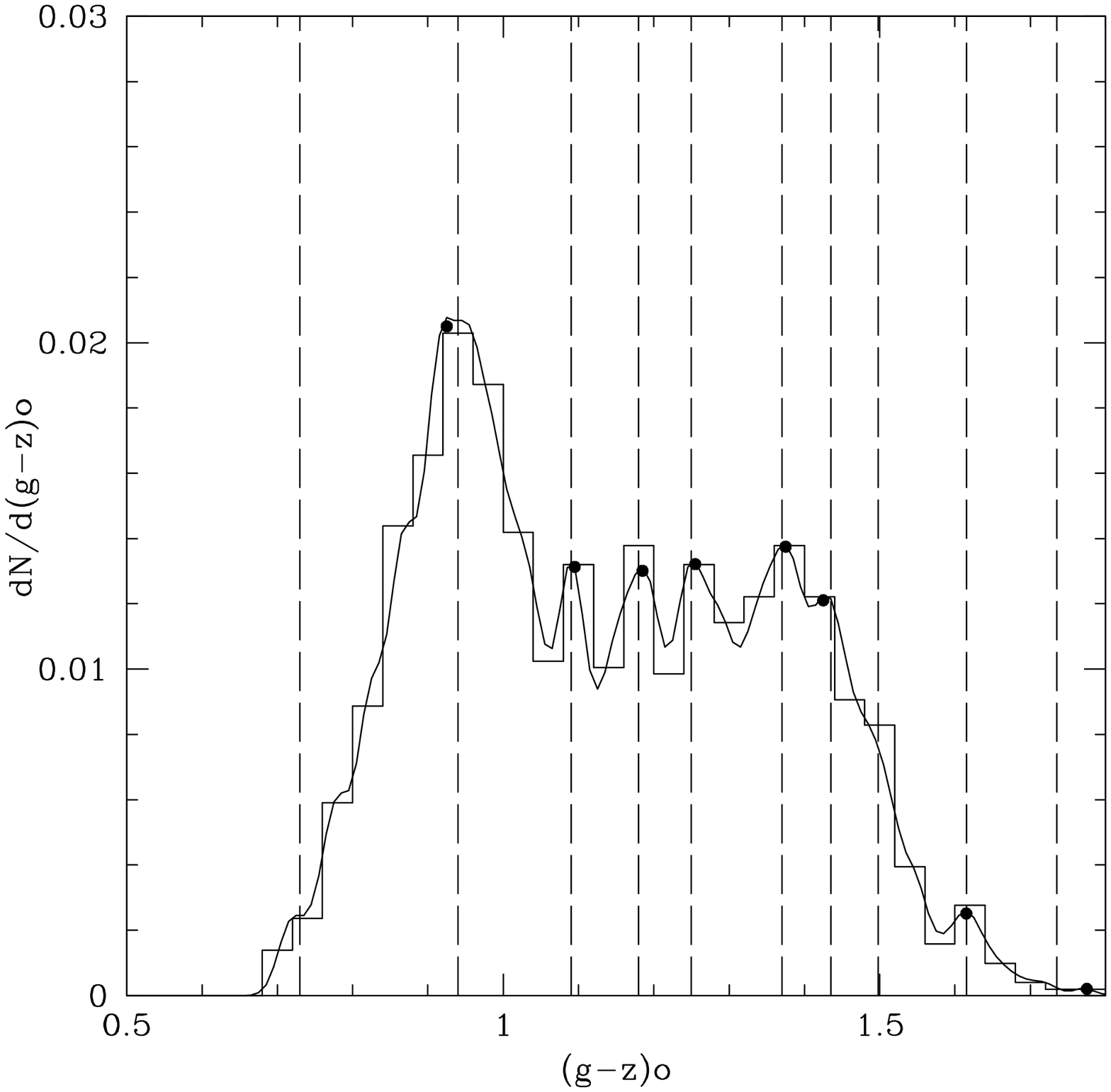}~\hfill
  \caption{Discrete and smoothed colour distribution for 1620 clusters in 7
Fornax galaxies with $M_{g}=$-20.2 to -19.2. The dashed lines indicate the Fornax
colour pattern.
}
  \label{fig:fig3}
\end{figure}

  The usual analysis of $GCCDs$ stands on discrete-bin colour histograms and/or
  smoothed versions of these histograms. These last "generalized" histograms
  are obtained by convolving the colour data with, for example, Gaussian kernels.
  In this work we adopt the same tools but, instead of sampling a single galaxy, we
  define a moving sampling window (0.4 mag. wide in galaxy absolute magnitude) and
  combine all the $GCs$ associated with galaxies within that window. In turn, this
  window moves in steps $\delta=$0.20 mag.\\
  A composite $GCCD$ is obtained for each galaxy sample, and a routine searches
  for colour peaks and valleys (i.e. colours where [dN/d(g-z)]=0). \\
  This procedure was carried out for all galaxies fainter than $M_{g}=$-20.2
  (i.e. non-giant galaxies),  leading to the identification of 231 peaks in Virgo
  and 179 in Fornax. The statistics of these peaks define a first approach to the 
  Virgo and Fornax colour patterns.\\
  A further analysis reveals that most of these patterns appear for $GCs$ associated
  to galaxies with absolute magnitudes $M_{g}$ from -20.2 to -19.2.\\
  The composite $GCCDs$ for 13 Virgo galaxies and 7 Fornax galaxies are
  shown (both in the discrete and smoothed histogram format) in Fig.~\ref{fig:fig2}
  and Fig.~\ref{fig:fig3} respectively. In these figures, the discrete-bin histograms have
  0.04 mag bins while, the smoothed distributions were obtained with a Gaussian
  kernel $\sigma_{(g-z)}=$ 0.015 mag. Dashed lines indicate the respective colour
  patterns and solid dots identify the features found by the peak finding routine.\\
  It is worth mentioning that colour patterns seen in these diagrams survive when the
  $GC$ samples are divided in terms of galaxy groups or in terms of the apparent magnitudes
  of the clusters.\\
  Within the $(g-z)$ colour range covered by old $GCs$ (0.75 to 1.65), the Virgo galaxies
  exhibit six (and possibly seven) colour peaks while five (and possibly 6) are detectable
  in Fornax galaxies. 
\section{Colour patterns in Virgo Giants.}
\begin{figure}[!ht]
  \centering
  \includegraphics[width=.5\textwidth]{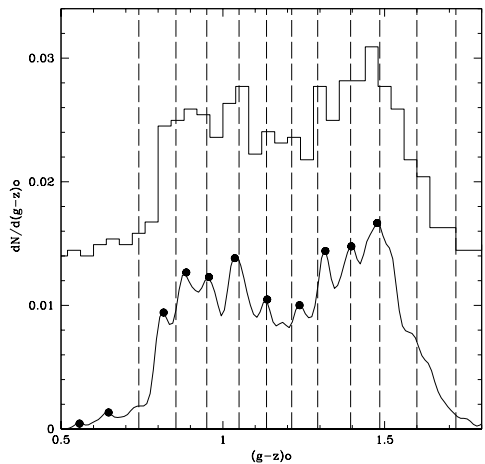}~\hfill
  \caption{$GC$ sample in $NGC~4486$. The sample includes 547 clusters with galactocentric radii
 from 0 to 110 arcsecs and position angles between 20 and 160 degrees. The dashed lines indicate
the Virgo colour pattern.
}
  \label{fig:fig4}
\end{figure}

\begin{figure}[!ht]
  \centering
  \includegraphics[width=.6\textwidth]{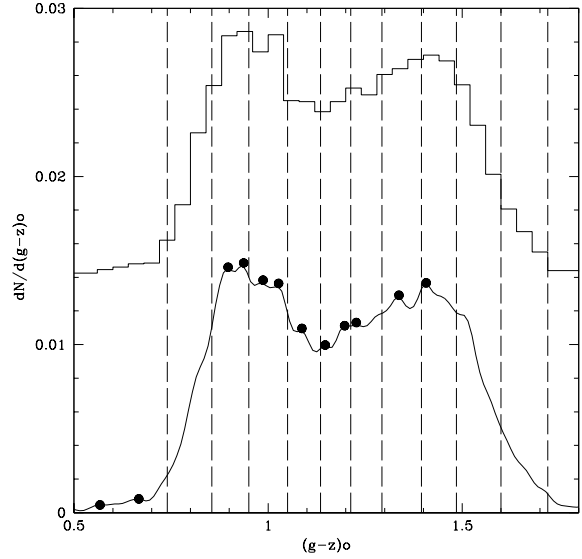}~\hfill
  \caption{Composite GC colour distribution for 4974 clusters in ten giant Virgo galaxies.
 The dashed lines indicate the Virgo colour pattern. Dots indicate the position of (incoherent)
 colour peaks identified by the finding routine. 
}
  \label{fig:fig5}
\end{figure}

\begin{figure}[!ht]
  \centering
  \includegraphics[width=.5\textwidth]{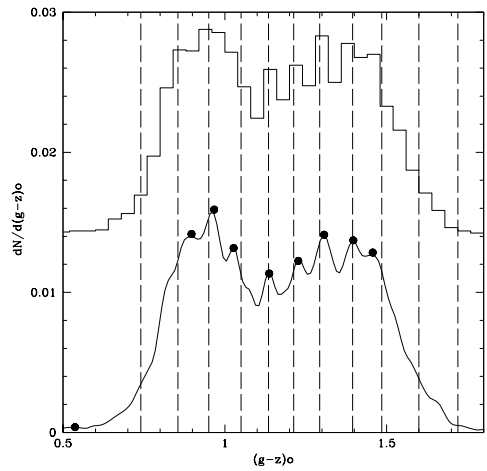}~\hfill
  \caption{Composite GC colour distribution for 4974 clusters in ten giant Virgo galaxies.
 The dashed lines indicate the Virgo colour pattern.
 The individual colour patterns for $GCs$ in each galaxy were shifted according to the results
 derived with the peak finding routine and then added (see text). The Virgo colour pattern is
 easily recognizable on the composite $GCCD$.
}
  \label{fig:fig6}
\end{figure}

 In this, and in the following section, we present the results for the giant
 galaxies in both clusters. In what follows, all the diagrams display the smoothed 
 $GC$ colour distributions and the discrete-bins histograms (arbitrarily shifted upwards).\\
 A remarkable object in Virgo is the giant galaxy $NGC~4486$ which has an extremely rich $GC$ system
 (with about 15.000 clusters). In this case the routine searches for colour patterns
 within given ranges of galactocentric radii and position angles. These patterns are
 then compared with the corresponding reference pattern (Virgo or Fornax).\\
 An example of the results for $NGC~4486$ is displayed in Fig.~\ref{fig:fig4}. This diagram shows eight
 coincidences with the colour pattern defined by $GCs$ in fainter galaxies, without requiring 
 any systematic shift in colour.\\
 In fact, the routine finds the Virgo pattern in all the ten giant galaxies shown in Fig.~\ref{fig:fig1}  although,
 in most cases, colour shifts ranging from -0.05 to 0.01 mag. are required for 
 a proper match with the reference pattern.\\
 The composite $GCCD$ for these giant galaxies is shown in Fig.~\ref{fig:fig5}. The colour
 distribution is broadly bimodal and the finding routine delivers a number of
 rather "incoherent" colour peaks (shown as dots). However, if the $GCCD$ in each galaxy
 is shifted in colour, as indicated by the routine, and then added, the Virgo pattern
 appears clearly defined as depicted in Fig.~\ref{fig:fig6}.
\begin{figure}[!ht]
  \centering
  \includegraphics[width=.5\textwidth]{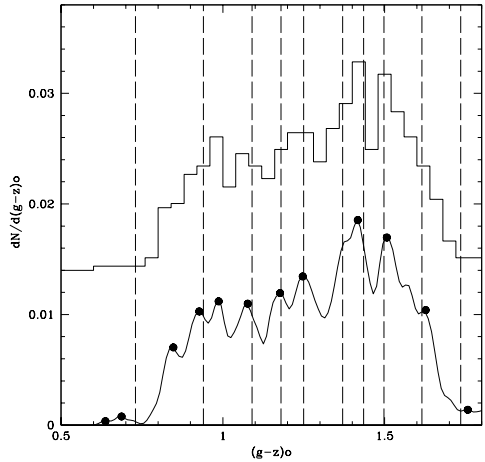}~\hfill
  \caption{$GC$ sample in $NGC~1399$. The sample includes 663 clusters with galactocentric radii
 from 0 to 90 arcsecs and position angles between 0 and 360 degrees. The dashed lines indicate
 the Fornax colour pattern.
}
  \label{fig:fig7}
\end{figure}
\begin{figure}[!ht]
  \centering
  \includegraphics[width=.5\textwidth]{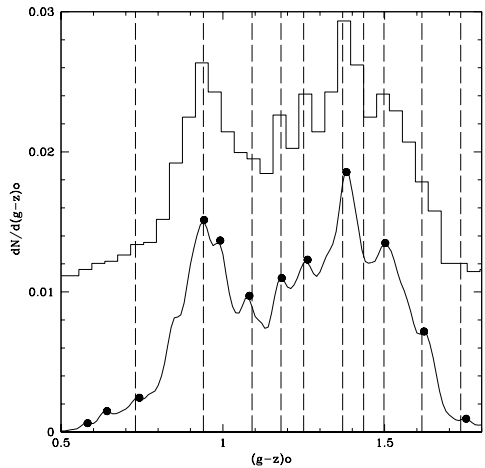}~\hfill
  \caption{Composite GC colour distribution for 1677 clusters in four giant Fornax galaxies, shifted
  by 0.035 in the $(g-z)$ colour. The dashed lines indicate the Fornax colour pattern.
}
  \label{fig:fig8}
\end{figure}
\begin{figure}[!ht]
  \centering
  \includegraphics[width=.5\textwidth]{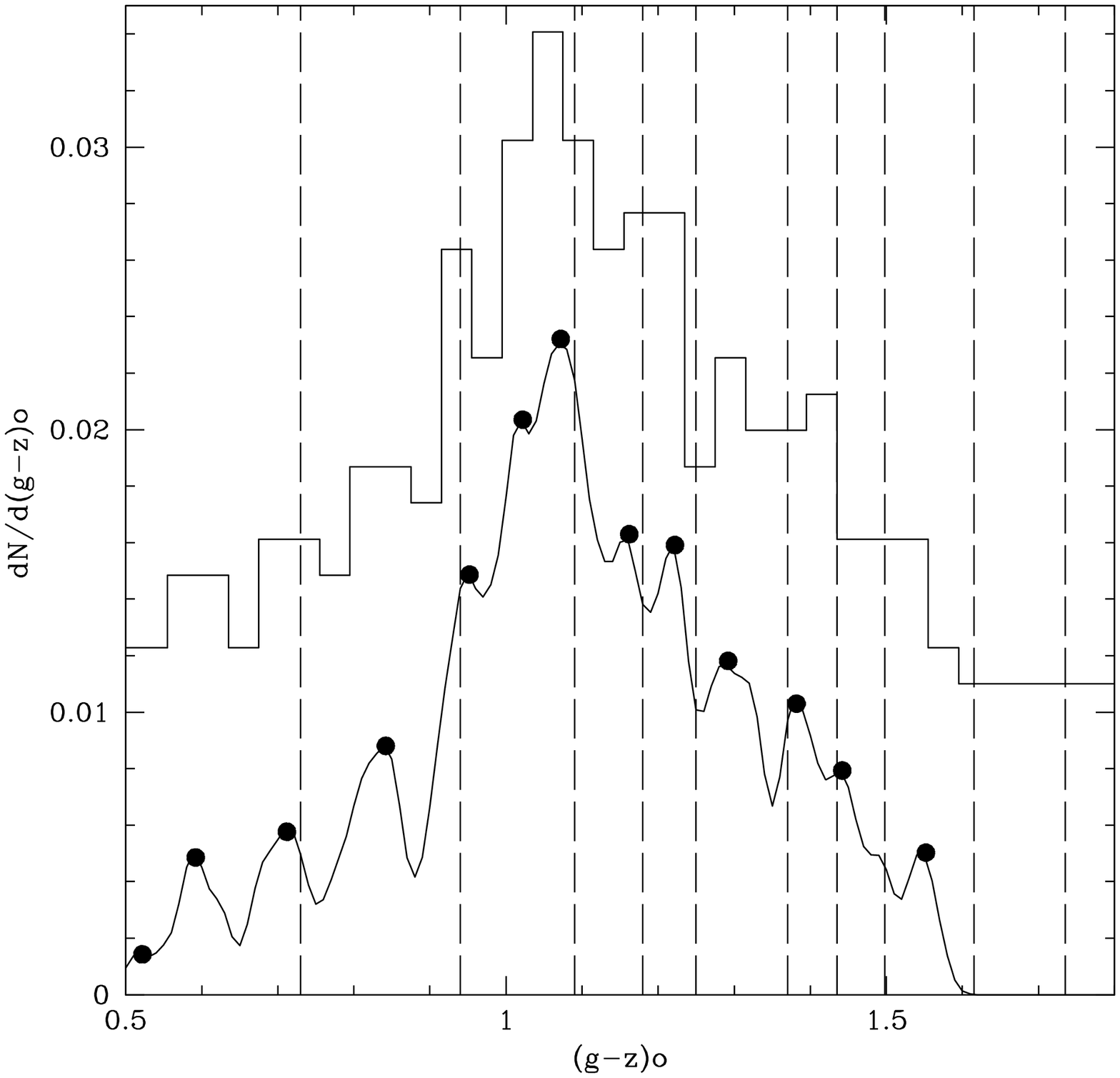}~\hfill
  \caption{Composite GC colour distribution for 239 clusters in $NGC~1316$, shifted
  by 0.035 in the $(g-z)$ colour. These clusters have galactocentric radii between 40
  and 80 arcsecs. The dashed lines indicate the Fornax colour pattern.
}
  \label{fig:fig9}
\end{figure}
\section{Colour patterns in Fornax Giants.}
 In the case of the Fornax cluster, the dominant elliptical galaxy is $NGC~1399$. In this
 galaxy the routine indicates a colour shift of 0.035  in order to match the
 Fornax pattern as displayed in Fig.~\ref{fig:fig7}. Furthermore, the same shift is 
 necessary when $GCs$ in the four brightest giants in Fornax are combined. This result
 is shown in Fig.~\ref{fig:fig8}. This diagram shows seven out of eight
possible coincidences with the Fornax pattern.\\
 This last diagram does not includes $NGC~1316$, a galaxy
 with a complex multi-population $GC$ system (see Sesto, Faifer and Forte, 2017). However,
 even in this galaxy, some of the colour peaks in the Fornax pattern can be recognized as
 depicted in Fig.~\ref{fig:fig9}.\\
 It is worth mentioning that the Washington $(C-T_{1})$ colours of clusters
 in $NGC~1399$ and $NGC~4486$ (Forte et al. 2007) show the same features detected
 on the $ACS$ photometry.\\
 The Virgo pattern is also recognizable in a peripheral field of $NGC~4486$ observed
 with $Gemini-GMOS$ and includes some 500 $GC$ candidates. (Forte et al 2013).\\
\section{Conclusions.}
 A remarkable outcome in this work is that, once the colour patterns where recognized
 on the composite $GCCDs$ of clusters associated with galaxies fainter than $M_{g}=$-20.2,
 the same features were later found in the $\it individual$ $GCCDs$ of giant galaxies
 both in Virgo and Fornax. This fact supports the physical entity of the colour patterns
 and argues against the idea that they are mere statistical fluctuations.\\
 The origin of these colour patterns is intriguing and, at this stage, only speculative.
 There are different scenarios that deserve further research in order to clarify this
 situation. Among them, the effects of $AGN$ activity or those connected with violent
 star formation events (e.g. arising in the merging of sub-cluster structures at high
 red shifts).\\
 The clusters arising in these events may co-exist with other $GCs$ that were
 formed along the individual life of a given galaxy. The rich $GCs$ in giant galaxies
 may in fact hide the presence of those clusters although they are still detectable
 through the pattern recognition technique.\\
 For galaxies fainter than $M_{g}=$-18.8, the colour patterns are not easily recognizable,
 except in a few cases. These systems do not have a significant population of intermediate
 colours and red $GCs$ to allow the eventual presence of the colour patterns.\\ 
 If chemical abundance correlates with time, as in the case of $MW$ globulars 
 (see Leaman, 2013), the different colour peaks may be indicating the time of the
 occurrence of "outer stimuli" that led to the enhancement of $GC$ formation
 on $Mpc$ spatial scales.\\
 The adoption of the $MW$ age-chemical abundance relation, just as a 
 reference (and with all the well known caveats), suggests a time lapse of about 1.5 $gy$, at
 red shifts $z$ between 2 and 4, as the ages of those putative highly energetic events.
 Further characterization of these patterns in an astrophysical context will require high
 quality photometry and spectroscopy.\\ 
\agradecimientos
 The author thanks the organizing Committees for the invitation to deliver this talk and
 participate in this meeting honouring a friend and outstanding colleague as Dr. Juan Jos\'e
 Clari\'a. C\'ordoba Observatory, Argentina, June 21-24, 2016.
\begin{referencias}
\vskip .5cm
\reference Cen, R., 2001, \apj 560, 592
\reference Fabian, A.C., 2012, \araa 50, 455
\reference Forte, J.C., Faifer, F.R., Geisler, D. \mnras 382, 1947
\reference Forte, J.C., et al., 2013, \mnras 431, 1405
\reference Jord\'an A. et al., 2009, \apjs 180, 54
\reference Jord\'an A. et al., 2015, \apjs 221, 13
\reference Leaman, R., VandenBerg D.A., Mendel, T., 2013, \mnras 436, 122
\reference Peng, E., et al., 2006, \apj 639, 838
\reference Sesto, L., Faifer, F.R., Forte, J.C., 2016 \mnras 461, 4260
\reference Silk, J., 2013, \mnras 461, 4260
\reference Vogelsberger, M. et al., 2014, \mnras 444, 1518
\end{referencias}
\end{document}